\documentclass[onecolumn]{aa}  

\usepackage{graphicx}
\usepackage{txfonts}
\usepackage[usenames,dvipsnames]{xcolor}
\usepackage{color}

\newcommand{\tn}{\textnormal}

\begin{document} 

\title{Constraints on the astrophysical environment of binaries\\
 with gravitational-wave observations}


   \author{Vitor Cardoso
          \inst{1}
          \and
          Andrea Maselli\inst{2}}

   \institute{CENTRA, Departamento de F\'{\i}sica, Instituto Superior T\'ecnico -- IST, Universidade 
   de Lisboa -- UL, Avenida Rovisco Pais 1, 1049 Lisboa, Portugal\\
              \email{vitor.cardoso@tecnico.ulisboa.pt}, 
         \and
             Dipartimento di Fisica, ``Sapienza'' Universit\`a di Roma \& Sezione 
             INFN Roma1, P.A. Moro 5, 00185, Roma, Italy\\
             \email{andrea.maselli@roma1.infn.it}
             }


 
  \abstract
   {}
   {The dynamics of coalescing compact binaries can be affected by the environment in which the systems evolve, 
   leaving detectable signatures into the emitted gravitational signal. In this paper we investigate the ability of 
    gravitational-wave detectors to constrain the nature of the environment 
   in which compact binaries merge.}
   {We parametrize a variety of environmental effects by modifying the phase of the gravitational signal 
emitted by black hole and neutron star binaries. We infer the bounds on such effects 
by current and future generation of interferometers, studying their dependence on the binary's parameters.}
   {We show that the strong dephasing induced by accretion and dynamical friction can constraint the 
density of the surrounding medium to orders of magnitude below that of accretion disks. Planned detectors, such 
as LISA or DECIGO, will be able to probe densities typical of those of dark matter}
   {}


   \maketitle
%

\section{Introduction}

Most of the content of our universe is unknown, and its 
properties may change from the solar neighborhood to 
distant galaxies. Accordingly, the astrophysical environment 
around stellar and massive black holes (BHs) can be very 
diverse. The dark matter density in the Solar neighborhood 
is of order $\rho=0.01M_{\odot}/pc^3=6.7\times 10^{-22}\,{\rm Kg}/{\rm m}^3$~\citep{Pato:2015dua} and that of interstellar dust 
can be even lower. However, the dark matter density can 
be 8 orders of magnitude (or more) larger, close to the 
center of galaxies and in the vicinities of BHs~\citep{Ferrer:2017xwm}. Supermassive BH binaries can be evolving in accretion 
disks which have baryonic densities as large as 
$10^{-6} - 100\,{\rm Kg}/{\rm m}^3$ for thick and thin 
accretion disks, respectively~\citep{Barausse:2014tra}. 
It has also been conjectured that coalescing BHs may form 
via dynamical fragmentation of a very  massive  star 
undergoing gravitational collapse, leading to a binary 
evolving in a medium with density as high as 
$10^{10}\,{\rm Kg}/{\rm m}^3$ or higher~\citep{Loeb:2016fzn,Reisswig:2013sqa}~\footnote{However, the lack of any counterparts for the 
LIGO-Virgo binary BH sources is evidence that such extreme 
scenarios are not common.}.

In the presence of a nontrivial environment (magnetic fields, 
fluids, dark matter, etc), three mechanisms contribute to 
change the dynamics of a compact binary with respect to 
that in vacuum: accretion, gravitational drag and the 
self-gravity of the fluid. These all contribute to a small, 
but potentially observable change of the gravitational-wave 
(GW) phase. We wish to understand to which level can GW 
observations constrain the properties of the environment, 
with only mild assumptions.

\section{The phase dependence in vacuum and beyond}
\subsection{Setup}

Take a BH binary of total mass $M=m_1+m_2$, with $m_{1,2}$ component masses
separated by a distance $L$ and with an orbital frequency $\Omega$.
Although the orbit is generically eccentric, GW emission tends to circularize it on relatively short timescales~\cite{Peters:1964zz,Krolak:1987ofj}.
This is certainly true for stellar mass BH binaries, which form substantially prior to merger and evolve mostly only via GW emission.
However, the formation of supermassive BH binaries is poorly understood. Some of the mechanisms that contribute to such binaries forming and merging actually may also impart a substantial eccentricity,
specially in its initial stages~\cite{Barack:2018yly}. However, even in such scenario the evolution of such systems quickly reduces eccentricity~\cite{Key:2010tc}.
Here, we will always assume that the binary circularized by the time it enters the detector band.
At leading order in vacuum GR, the dynamics of a binary is governed by energy balance:
the quadrupole formula implies that the binary is emitting a flux of GW energy
\begin{equation}
\dot E_{\rm GW}=\frac{32}{5}\mu^2 L^4\Omega^6\,, \label{quadrupole}
\end{equation}
in GWs, where $\mu=m_1 m_2/M$ is the reduced mass of the system. Thus, the orbital energy of 
the system $E_{\rm orb}=-M\mu/(2L)$ must decrease at a rate fixed by such loss. This fixes 
immediately the time-dependence of the GW frequency to be~$f^{-8/3}=(8\pi)^{8/3}{\cal M}^{5/3}(t_0-t)/5$, 
where ${\cal M}=M\eta^{3/5}$ is the chirp mass, $\eta=m_1 m_2/M^2= \mu/M$ the symmetric mass ratio,
and $f=\Omega/\pi$.

Once the frequency evolution is known, the GW phase simply reads
\begin{equation}
\varphi(t)=2\int^t\Omega(t')dt' \,.\label{GWphase}
\end{equation}

In Fourier domain it is possible to obtain analytical templates of the waveforms. 
One can write the metric fluctuations as
\begin{eqnarray}
 h_+(t)&=&A_+(t_{\rm ret})\cos\varphi(t_{\rm ret}) \,,\\
 h_\times(t)&=&A_\times(t_{\rm ret}) \sin\varphi(t_{\rm ret})\,,
\end{eqnarray}
where $t_{\rm ret}$ is the retarded time. The Fourier-transformed quantities are
\begin{equation}
 \tilde{h}_+= {\cal A}_+e^{i\Psi_+}\,,\qquad  \tilde{h}_\times={\cal A}_\times e^{i\Psi_\times}\,.
\end{equation}

Dissipative effects are included within the stationary phase approximation, where the secular time evolution is governed by 
the GW emission \citep{Flanagan:1997sx}. In Fourier space, we decompose the phase of the GW signal 
$\tilde{h}(f)={\cal A}e^{i\Psi(f)}$ as:  
\begin{equation}
 \Psi(f)
 =\Psi_{\rm GR}^{(0)}[1+{\rm (PN\ corrections)}+\delta_{\Psi_{\rm env}}]\,.
\end{equation}
where $\Psi_{\rm GR}^{(0)}=3/128 ({\cal M}\pi f)^{-5/3}$ represents the leading term of the 
phase's post-Newtonian expansion.

\subsection{Corrections to the GW phase due to environmental effects}

%
\begin{table}
\caption{Corrections $\delta_{\Psi_{\rm env}}=\kappa_{\beta} \rho_0 M^{2-\beta}R^\beta (M f)^{-\gamma}$ to the GW phase in the Fourier space computed 
within the stationary phase approximation for a quasicircular binary (see Eq.~(\ref{eq:final})). The binary moves in a medium of 
density $\rho=\rho_0 (R/r)^{\beta}$, and is subjected to accretion, gravitational forces from the matter distribution (assumed to be 
centred at the binary's center of mass) and to gravitational drag. Results refer to $\beta=(0,1)$ and to collisionless and Bondi accretion, 
respectively.}
\begin{tabular}{c|c c| cc}
 \hline\hline
Mechanism                                 & $\gamma_{(\beta=0)}$& $\kappa_{0}$                        & $\gamma_{(\beta=1)}$&$\kappa_1$\\
\hline
Gravitational pull                        &    2                & $1$                                  & 4/3                  & 1\\
Gravitational drag                        & 11/3                & $-\eta^{-3}(1-3\eta)\pi^{-11/3}$  &3                    &$-\eta^{-16/5}(1-3\eta)\pi^{-3}$\\
accretion - Bondi                         & 11/3                & $-\eta^{-3}(1-3\eta)\pi^{-11/3}$  & 3	                   &$-\eta^{-16/5}(1-3\eta)\pi^{-3}$\\
accretion - collisionless                 & 3                 &$-\eta^{-1}\pi^{-3}$                 & 7/3                 &$-\eta^{-6/5}\pi^{-7/3}$	\\
\hline\hline
\end{tabular}
\label{tab:dephasing_fourier}
\end{table}
Environmental effects in the evolution of a compact binary can be divided in different categories, and were comprehensibly 
studied in the past~\citep{Yunes:2011ws,Kocsis:2011dr,Eda:2013gg,Macedo:2013qea,Barausse:2014tra}. We summarize here the results of~\citep{Barausse:2014tra}
extending them to generic density profiles. Consider a BH binary evolving in a medium of density
\begin{equation}
\rho=\rho_0(R/r)^{\beta}\,.\label{powerlawdensity}
\end{equation}
This density profile can describe a constant magnetic field or constant-density fluid for $\beta=0$, or thick accretion disks.
Studies of fuzzy dark matter, described by ultralight scalars, show that close to the galactic centers the density is approximately constant, 
$\beta\approx 0$~\citep{Hui:2016ltb,Annulli:2020ilw}.
On the other hand, particle-like dark matter (with small Compton wavelength) is described
by $\beta=1$ in the inner core of dark matter regions~\citep{Navarro:1996gj,Gondolo:1999ef,Sadeghian:2013laa}. However, such large overdensities are easily washed out
via scattering of stars or BHs, or accretion by the central BH, induced by heating in its
vicinities~\citep{Merritt:2002vj,Bertone:2005hw,Merritt:2003qk}. These effects tend to 
smooth the dark matter density close to the center of galaxies. We thus take $\beta=0$ to describe well most of the known environment around compact binaries.
We note that an arbitrary density profile is locally constant, and therefore $\beta\approx 0$ for the physics of {\it compact} binaries, the focus of this work.

Such an environment affects the binary dynamics in different ways: by exerting a gravitational pull on the binary,
the Newtonian equation of motion and balance equations change, leading to a relative dephasing
\begin{equation}
\delta_{\Psi_{\rm env}}^{\rm grav\, pull}\approx\rho_0 f^{\frac{2\beta}{3}-2} M^{-\beta/3}  R^{\beta} \,,\label{psi_grav_pull}
\end{equation}
up to factors of order unity, which agrees with previous results~\citep{Eda:2013gg,Barausse:2014tra}. 

Accretion of the surrounding medium into the BH also introduces a dephasing. This can be estimated by assuming a fluid at rest (with 
no angular momentum) and unperturbed by the binary. The dephasing introduced by accretion can be computed by extending previous analysis~\citep{Macedo:2013qea,Barausse:2014tra}. 
The result depends on the type of accretion (i.e., if the environmental medium is 
collisionless or behaves as a fluid). We find
\begin{equation}
 \delta_{\Psi_{\rm env}}^{\rm accretion}\approx \left\{\begin{array}{c}
                   -\frac{R^{\beta}\rho_0}{\eta^{2/5}{\cal M}^{1+\frac{\beta}{3}}}(\pi  f)^{\frac{2\beta}{3}-3} \quad {\rm collisionless}\\
                   -\frac{(1-3\eta) R^{\beta}\rho_0 }{\eta^2{\cal M}^{\frac{5}{3}+\frac{\beta}{3}}} (\pi  f)^{\frac{2\beta}{3}-\frac{11}{3}} \quad {\rm Bondi}
                  \end{array}\right.\label{phase_accretion}
\end{equation}

Besides the extra gravitational pull by the matter inside the orbital radius and accretion, all of the surrounding medium slows the binary down through gravitational drag. At leading order, the gravitational drag produces a force on object ``i'' in the direction of motion, given by
\begin{equation}
F_{\rm DF}= \frac{4\pi\rho (G M_i)^2}{v_i^2} I\,,
\end{equation}
where $v$ is the relative velocity between body ``i'' and the gas (which we will take, as a rough approximation, to be given by the Keplerian velocity), and $I$ is a prefactor of order unity which depends on the relative velocity of the binary components with respect to the medium~\citep{Kim:2007zb}.

For supersonic motion, $I \sim \ln\left[(r_i/r_{\min})/(0.11\Upsilon+1.65)\right]$, with $\Upsilon\equiv v/v_s$ the Mach number ($v_s$ is the sound speed), $r_i$ the orbital radius of the object and $r_{\min}$ an unknown fitting parameter. Our results show a very mild dependence on the exact value of $r_{\min}$. There are special-relativistic corrections to this formula~\citep{Barausse:2007ph}, and corrections for slab-like geometries (such as those in accretion disks)~\citep{Vicente:2019ilr}. We do not consider these effects here. Relativistic effects are expected to have a negligible impact: dynamical friction on a straight-moving object is affected by a correction term of the order $(1+v^2)^2\Gamma^2$, with $\Gamma$ the Lorentz factor and $v$ the relative velocity~\citep{Barausse:2007ph}. Even for $v=0.1$, this introduces a correction of only $22\%$ in the dynamical friction, not affecting our order-of-magnitude estimates. The geometry of the environment, on the other hand, may be relevant for thin accretion disks. When the motion is supersonic, geometry may suppress friction by a factor two~\citep{Vicente:2019ilr}. Suppression is stronger for subsonic motion.

The dephasing introduced by dynamical friction is regulated by an extra energy loss
$\dot{E}_{\rm DF}= F_{\rm DF} v_K$.
For the profile (\ref{powerlawdensity}), we find the dephasing
\begin{equation}
 \delta_{\Psi_{\rm env}}^{\rm DF}\approx -\frac{(1-3\eta) R^{\beta} \rho_0}{\eta^2{\cal M}^{\frac{1}{3} (\beta+5)}}(\pi  f)^{\frac{2\beta}{3}-\frac{11}{3}}\,. \label{phase_drag}
\end{equation}

The results for the dephasing can all be captured by the expression
\begin{equation}
\delta_{\Psi_{\rm env}}=\kappa_{\beta}M^{2-\beta}R^\beta\rho_0 (Mf)^{-\gamma}\label{eq:final}\,,
\end{equation}
up to factors of order unit, where the exponent $\gamma$ and the precise coefficient $\kappa_\beta$ are listed in Table~\ref{tab:dephasing_fourier}. 
The parameter $\beta$ only affects the results via a simple re-scaling of the $\beta=0$ results. 
In the rest of this work we focus on $\beta=0$: in this case the most  
constraining effect is given by the gravitational drag and it is controlled by $\gamma=11/3$.

Finally, we note that the corrections introduced by environmental effects modify the phasing at very low (and 
negative!) {\it pre}-Newtonian order. They can all be included in a parametrized formalism~\citep{Yunes:2009ke,Barausse:2014tra,TheLIGOScientific:2016src}. Some of 
these modifications of the PN series were already considered in the context of extra radiation 
channels, with an unknown underlying physical theory~\citep{Barausse:2016eii,Carson:2019fxr,Gnocchi:2019jzp}. 
The results above show that such modifications appear at $n=-3\gamma/2$ PN order, and have 
a very specific physical origin. 

Our purpose now is to constrain, at an order-of-magnitude level and agnostically, the environmental properties. Although the drag created by thin accretion disks falls outside the approximations made here for the density, it can be mapped into one of the above with $\beta=15/8$~\citep{Barausse:2014tra}.

\subsection{Measurement uncertainties: a Fisher matrix analysis}
A comprehensive survey of the impact of several environmental effects in the GW signal is shown in~\citep{Barausse:2014tra},
through an estimate of the signal dephasing with respect to that of vacuum GR.
Here, we wish to quantify the precision with which current and planned detectors are able to constraint environmental properties.
As a baseline for the GR waveform we use the semi analytical PhenomB\footnote{The waveform corrections 
described in the previous section belong to the class of ``pre-Newtonian'' modifications, and they affect the 
signal at very low frequencies, where PhenomB is indistinguishable from more sophisticated 
templates \citep{Yunes:2009hc}.} template in the frequency domain for non-precessing spinning 
BHs \citep{Ajith:2007kx,Ajith:2009bn}. We consider a Newtonian amplitude, averaging on the sky localization of the binary systems. In the 
limit of GW signals with large signal-to-noise ratio, the probability distribution of the source's parameters, for a given observation, can be 
described by a multivariate Gaussian peaked around the true values, and with covariance $\Sigma_{ij}=(\Gamma^{-1})_{ij}$, given by the 
inverse of the Fisher matrix \citep{Vallisneri:2007ev}. The latter is build from the first derivatives of the GW template 
$\tilde{h}(f)={\cal A}\exp[i(\Psi_{\rm GR}+\psi_{\rm env})]$ with respect to the source's parameters 
$\theta_i=(\ln {\cal M},\ln \eta,\tau_c,\phi_c,\chi_{\rm eff},\rho_0)$:
\begin{equation}
\Gamma_{ij}=\int_{f_{\rm min}}^{f_{\rm max}}\frac{1}{S_n(f)}\frac{\partial \tilde{h}}{\partial \theta_i}\frac{\partial \tilde{h}}{\partial \theta_j}df\ . \label{fisher}
\end{equation}
The uncertainties on $\theta_i$ are given by the diagonal components of the covariance matrix, 
namely $\sigma_i=\sqrt{\Sigma_{ii}}$. Beside the chirp mass, the GW template depends 
on the symmetric mass ratio $\eta$, on the time and phase at the coalescence 
$(\tau_c,\phi_c)$, and on the effective spin $\chi_{\rm eff}=(m_1\chi_1+m_2\chi_2)/M$, where 
$\chi_{1,2}$ are the BH's dimensionless spin parameters. The integral in Eq.~(\ref{fisher}) 
is also function of the detector's noise spectral density $S_n(f)$. In our analysis we focus on 
both ground- and space-based interferometers. We consider advanced LIGO/Virgo at 
design sensitivity~\citep{LIGOWhite}, a third generation detector like the Einstein 
Telescope~\citep{Hild:2010id,ETWhite} (see also \cite{CosmicExplorer} for other proposals of 
third generation ground based detectors like the Cosmic Explorer), the LISA 
mission~\citep{2017arXiv170200786A}, and the Japanese satellite DECIGO, proposed to 
operate in the decihertz regime~\citep{Isoyama:2018rjb}. The lower end of the Fisher matrix's 
integration is set to $f_{\rm min}^{\rm LIGO/Virgo}=10$ Hz, $f_{\rm min}^{\rm ET}=3$Hz and 
$f_{\rm min}^{\rm DECIGO}=0.01$ Hz. For LISA we choose $f_{\rm min}^{\rm LISA}$ as min$[10^{-5}\tn{Hz},f_{obs}]$, where $f_{obs}$ is the 
frequency's value of the binary 4 years before the merger~\citep{Berti:2004bd}. On the other 
edge of the integral, we set the maximum frequency of ground based detectors to coincide with 
the PhenomB inspiral-merger transition value $f_{\rm 1M}$, which depends on the source's 
parameters~\citep{Ajith:2009bn}, while for space interferometers $f_{\rm max}^{\rm LISA}=\min[1\ {\rm Hz},f_{\rm 1M}]$ 
and $f_{\rm max}^{\rm DECIGO}=\min[100\ {\rm Hz},f_{\rm 1M}]$. 
Given the frequency content of the astrophysical corrections \eqref{psi_grav_pull}-\eqref{phase_accretion} 
and \eqref{phase_drag}, we expect  $f_{\rm max}$ to have a small effect on the uncertainties 
inferred through the Fisher matrix \citep{Barausse:2014tra}.

\section{Results: the ability of GW detectors to constrain environmental densities}

\begin{figure}[th]
\begin{tabular}{cc}
\includegraphics[width=8cm]{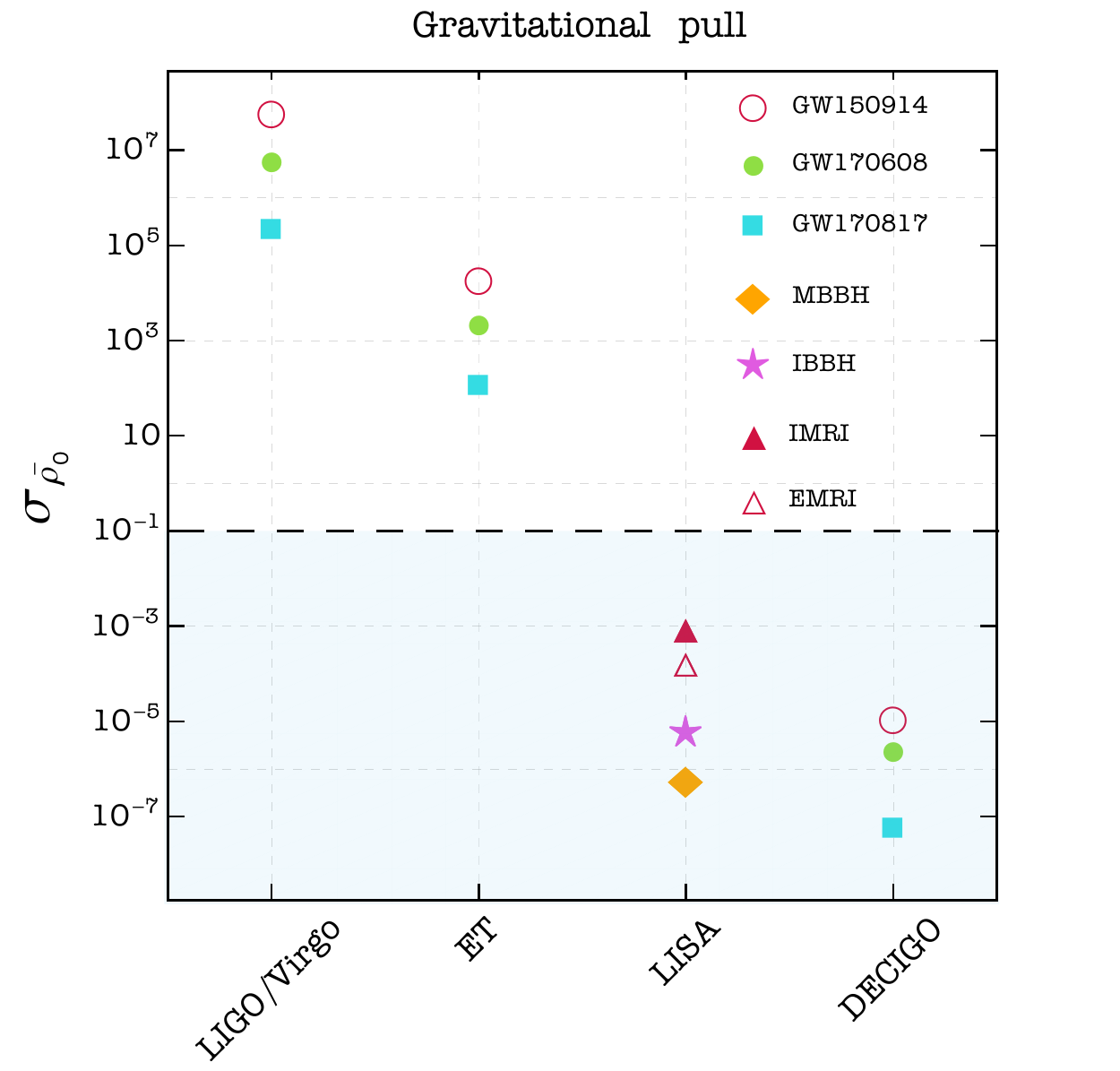}&
\includegraphics[width=8cm]{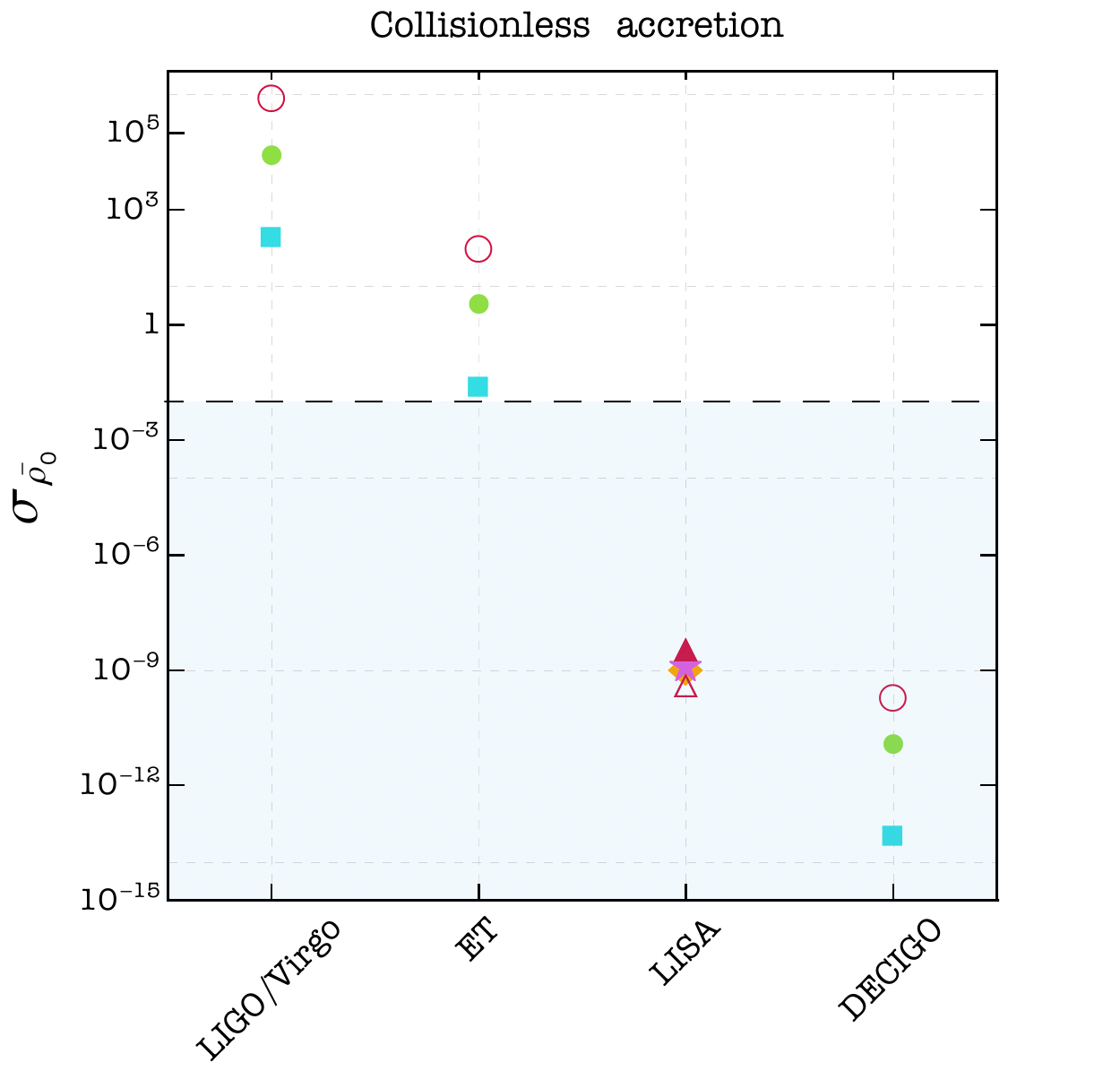}
\end{tabular}
\caption{1-$\sigma$ uncertainties on the density parameter $\rho_0$ normalized to the average 
density of water $\rho_{\rm H_2O}\simeq 10^3$kg/m$^3$ for different sources and detector 
configurations. Top and bottom panels refer to environmental effects due to gravitational 
pull and collisionless accretion. Different point markers identify distinct sources. For 
ground-based detectors and for DECIGO we consider sources with the same parameters 
of GW150914, GW170608 and GW170817~\citep{LIGOScientific:2018mvr}. For LISA we 
consider a massive and an intermediate-massive systems, as well as IMRI and EMRI, 
all located at $d=1$ Gpc from the detector.
The grey area denotes densities typical of accretion disks.}
\label{fig:constraints} 
\end{figure}
\begin{figure}[th]
\centering
\includegraphics[width=8cm]{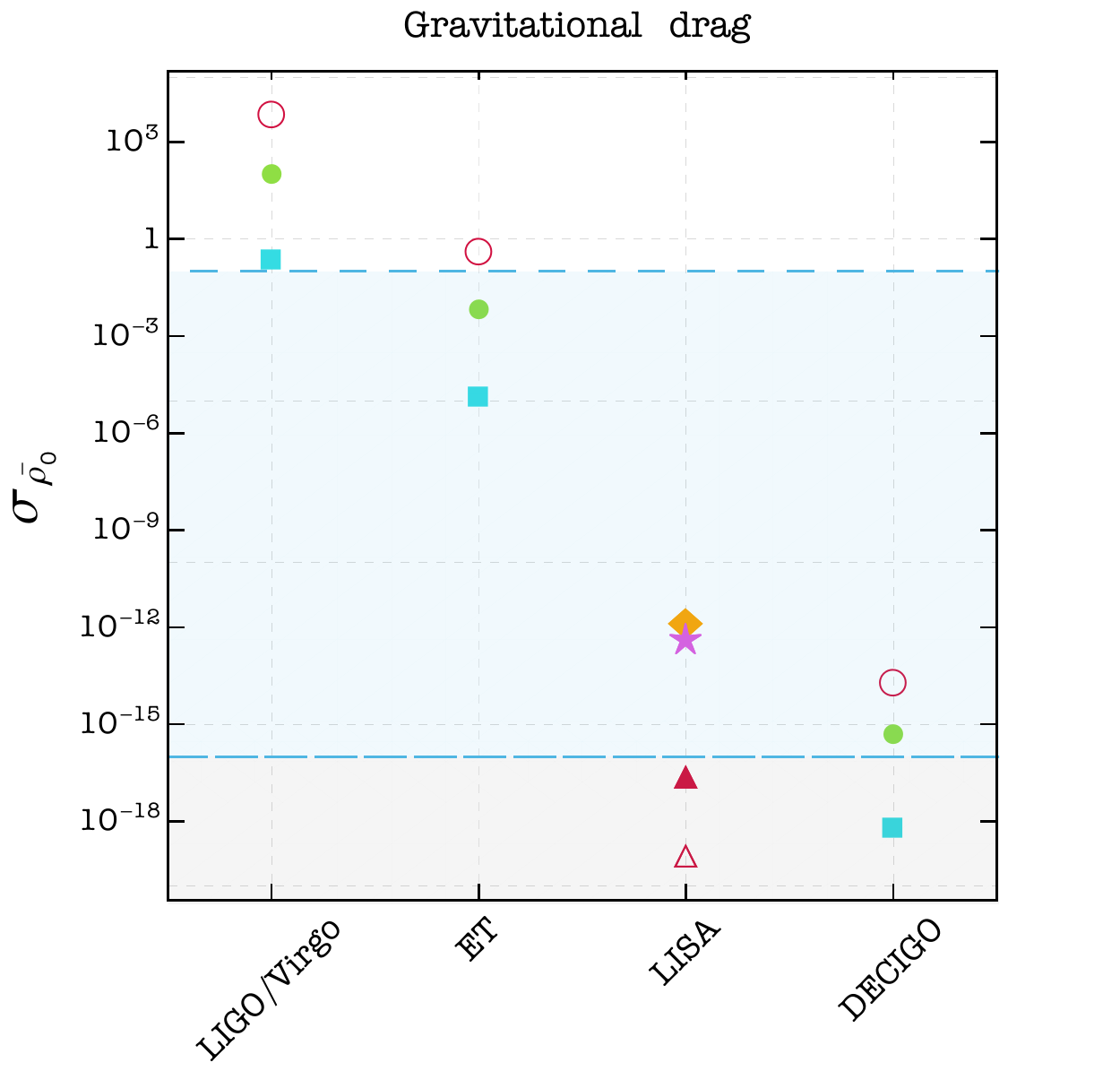}
\caption{1-$\sigma$ uncertainties on the density parameter $\rho_0$ normalized to the water average 
density for sources and detector configurations already shown in Fig.~\ref{fig:constraints}, obtained 
via (lack of) dephasing from dynamical friction. The shaded blue area denotes densities typical 
or smaller than that of accretion disks, while the gray shaded region denotes densities typical of 
dark matter.}
\label{fig:constraints2} 
\end{figure}
As prototype binary for our numerical analysis we consider five classes of 
objects: 

\noindent (i) two stellar mass BH binary systems with the properties of the observed 
gravitational events GW150915 and GW170608, and the binary neutron star event 
GW170817~\citep{LIGOScientific:2018mvr};

\noindent (ii) a massive binary (MBBH) with 
$(m_1,m_2)=(10^6,5\times 10^5)M_\odot$ and $(\chi_1,\chi_2)=(0.9,0.8)$;

\noindent (iii) 
an intermediate-mass binary (IBBH) with $(m_1,m_2)=(10^4,5\times 10^3)M_\odot$ 
and $(\chi_1,\chi_2)=(0.3,0.4)$;

\noindent (iv) an intermediate mass ratio inspiral (IMRI) 
where $(m_1,m_2)=(10^4,10)M_\odot$ and $(\chi_1,\chi_2)=(0.8,0.5)$;

\noindent (v) finally an extreme mass ratio binary (EMRI) with $(m_1,m_2)=(10^5,10)M_\odot$ and spin 
parameters $(\chi_1,\chi_2)=(0.8,0.5)$.

All massive sources targeted by LISA are located at 1 Gpc from the detector, 
while for the stellar binaries we use the median values of the luminosity distance 
estimated by the LIGO/Virgo collaboration. Depending on the interferometer and 
on the luminosity distance, the binaries feature different signal-to-noise ratios (SNR). 
For MBBH, IBBH, EMRIs and IMRIs detected by LISA we have SNR $\simeq 3\cdot10^4$,
$\simeq 708$, $\simeq 22$ and $\simeq 64$, respectively.  For the stellar mass systems 
in (i), (a) (SNR$^\tn{DEC}$,SNR$^\tn{ET}$,SNR$^\tn{LIGO})\simeq(2815,615,40)$ for GW150914; 
(b) (SNR$^\tn{DEC}$,SNR$^\tn{ET}$,SNR$^\tn{LIGO})\simeq(1290,303,21)$ for GW170608 and 
(c) (SNR$^\tn{DEC}$,SNR$^\tn{ET}$,SNR$^\tn{LIGO})\simeq(2124,502,35)$ for GW170817.

We use the Fisher matrix analysis above to study the lowest environmental densities that different GW detectors
are able to constrain. To be more specific, we study the critical density $\rho_0$ for which $\sigma_{\rho_0}\lesssim \rho_0$, the threshold for densities below which observations are unable to
distinguish between vacuum and a nontrivial environmental density.
Numerical values of the $1$-$\sigma$ uncertainties on the the parameter $\rho_0$ are shown in Figs.~\ref{fig:constraints}--\ref{fig:constraints2} 
for the three environmental effects discussed in the previous section.
We opted to normalize these results by the average density of the water $\rho_{\rm H_2O}\simeq 10^3$kg/m$^3$, defining
$\bar{\rho}_0=\rho/\rho_{\rm H_2O}$. This is an ad-hoc choice made only for visual clarity of the figures.

Figure~\ref{fig:constraints} shows that the gravitational pull of matter, even when optimized (we have centered the environment 
at the binary's center-of mass to maximize the effects of the pull) can provide only weak 
constraints. For a double neutron star, LIGO is able to bound the surrounding density to level $100,000$ smaller than that of water. The blue-shaded area in all figures indicates ranges of density of the order of those of accretion disks. As we can see, even using only the gravitational pull or accretion effects, future detectors will probe this region. Accretion effects are more important, and already the Einstein Telescope can probe environmental densities which can be of order of those 
expected for thin accretion disks.

As expected from the correspondence between the exponent $\gamma$ and the post-Newtonian order,
the most constraining effect is that of gravitational drag, which appears at $-5.5$ PN order. Although the data shown in 
Figs.~\ref{fig:constraints}-\ref{fig:constraints2} correspond to a specific binary location, our results can 
immediately be rescaled to any value of the source's luminosity distance, as the uncertainties are proportional to the 
inverse of $d$.

For stellar-mass binaries, the tightest constraints on the density $\rho_0$ are provided by DECIGO, 
yielding bounds several orders of magnitude stronger than those obtained with current and future generation of ground-based detectors. Constraints inferred by the Japanese 
satellite are also tighter than those derived from the observation of intermediate and massive sources in the millihertz regime by LISA. 
The best-case scenario is given by extreme/intermediate mass ratio inspirals detected by LISA, or by double neutron 
star system like GW170817, and in general by very light 
sources. 
The latter span a large number of cycles in the low-frequency part of the spectrum, where environmental 
effects are more important.

Results for the gravitational drag are particularly interesting. Advanced detectors may be able to constrain the parameter 
$\rho_0$ at values close to the typical densities of thin accretion disks. A third generation of detectors supplied by low-mass 
observations would also be able to probe densities featured by thick accretion disks. 
These values improve dramatically for space interferometers. Numerical data for DECIGO show that neutron star binaries 
are in principle able to constrain the lack of dephasing from dynamical friction at the level of dark matter density, namely 
for $\rho\ll 10^{-18}$Kg/m$^3$. Such results also extend to the case of Bondi accretion, for which the phase correction 
$\delta_{\Psi_{\rm env}}$ is characterized by the same exponent $\gamma$, and therefore by the same -5.5 PN order.

\section{Conclusions}

Gravitational wave observations of compact binary coalescences 
represent an established field of research, which is mapping the 
high-energy Universe detecting sources at various distances and 
orientations in the sky. Advanced detectors like LIGO/Virgo are now 
close to reach their full sensitivity, and will be soon joined by the 
Japanese KAGRA \citep{Akutsu:2018axf}. Moroever, third-generation 
ground-based detectors, as well as future space-based facilities, 
will make GW detections a weekly routine. The quantity and the quality 
of this incoming flood of data will allow to deepen our understanding 
on compact objects~\citep{Barack:2018yly,Cardoso:2019rvt}, and at the same time 
to investigate the arena in which they evolve. 

In this paper we have made a step forward in this direction, showing how 
GW detectors have a tremendous potential to constrain the properties of 
different phenomena occurring during the coalescence of binary sources. 
We have focused on three main effects, the gravitational drag, 
collisionless accretion and the gravitational pull, showing how observations 
translate into precise bounds on the binary environmental density.
We have investigated the precision of future measurements of this parameter 
by suitable modifications of the signal emitted by different families 
of compact objects. Our results are agnostic as to the nature of the 
environment, but can be easily mapped to specific models.
The bound inferred by space detectors as LISA and DECIGO are so tight 
that one can think on using GW detectors to exclude large dark matter 
overdensities close to the binary location.

The description given in this paper of dynamical friction holds for 
fluid-like environments, composed of particles with a small Compton 
wavelength. The calculation of dynamical friction for ultralight fields 
with a Compton wavelength large when compared to the binary parameters 
was completed after the completion of our work~\cite{Annulli:2020ilw}. 
This analysis predicts a dephasing $\delta_{\Psi_{\rm env}}$ described 
by Eq.~\eqref{eq:final} with $\gamma=4$. We have repeated the analysis 
described in the main text for such value of $\gamma$, finding 
that the detection of a IBBH with LISA will improve constraints on 
the density by over an order of magnitude. 

Our results assume that systematics -- such as uncertainties in computing the 
vacuum waveforms -- are under control.  In this context, we have tested our calculations by varying the template used for the Fisher analysis. This was done 
by (i) neglecting for example the contribution of the late inspiral/merger phase, 
and (ii) adding new parameters that can model missing physical effects, like 
tidal interactions. The uncertainties on the environmental parameters are 
robust against these changes, with relative differences of the order of 
sub-percent. The only exception is given by EMRIs/IMRIs for which for case 
(i) we find larger deviations on $\sigma_{\bar{\rho}_0}$, still $\lesssim 40\%$. Overall these changes do not alter the conclusions of our analysis. 
Our results are, naturally, a source of degeneracy and a limiting factor for 
tests of gravity~\citep{Barausse:2014tra,Yunes:2016jcc}: some of these 
astrophysical effects might be dominant over possible modifications of 
general relativity. The estimates of the leading post-Newtonian effects on 
the waveforms are simplistic, done in a Newtonian setup and neglecting 
backreaction on the medium itself. For equal-mass mergers this may mean 
that our drag estimates overestimate the actual effect.
Dynamical friction was handled at the non-relativistic level and neglecting 
effects of the geometry of the environment. Such effects may be important 
for thin accretion disks, for example, but the generalization of the results 
of \citep{Vicente:2019ilr} to circular motion is still missing. 
Possible multi-band detections were not explored here, but they will improve 
the bound we discussed even further. Our results highlight the amazing 
possibilities of GW astronomy, but also highlight the need to understand in 
detail the evolution of massive binaries within nontrivial environments.

\begin{acknowledgements}
V.C. acknowledges financial support provided under the European Union's H2020 ERC 
Consolidator Grant ``Matter and strong-field gravity: New frontiers in Einstein's 
theory'' grant agreement no. MaGRaTh--646597. A.M acknowledges support from the Amaldi 
Research Center funded by the MIUR program ``Dipartimento di Eccellenza'' (CUP: B81I18001170001).
This project has received funding from the European Union's Horizon 2020 research and innovation 
programme under the Marie Sklodowska-Curie grant agreement No 690904.
We acknowledge financial support provided by FCT/Portugal through grant PTDC/MAT-APL/30043/2017.
The authors would like to acknowledge networking support by the GWverse COST Action 
CA16104, ``Black holes, gravitational waves and fundamental physics.''
\end{acknowledgements}

\bibliographystyle{aa} 
\bibliography{References}

\begin{thebibliography}{45}
\expandafter\ifx\csname natexlab\endcsname\relax\def\natexlab#1{#1}\fi

\bibitem[{Abbott {et~al.}(2016)}]{TheLIGOScientific:2016src}
Abbott, B.~P. {et~al.} 2016, Phys. Rev. Lett., 116, 221101, [Erratum: Phys.
  Rev. Lett.121,no.12,129902(2018)]

\bibitem[{Abbott {et~al.}(2017)}]{CosmicExplorer}
Abbott, B.~P. {et~al.} 2017, Class. Quant. Grav., 34, 044001

\bibitem[{Abbott {et~al.}(2018)}]{LIGOScientific:2018mvr}
Abbott, B.~P. {et~al.} 2018 [\eprint[arXiv]{1811.12907}]

\bibitem[{Ajith {et~al.}(2008)}]{Ajith:2007kx}
Ajith, P. {et~al.} 2008, Phys. Rev., D77, 104017, [Erratum: Phys.
  Rev.D79,129901(2009)]

\bibitem[{Ajith {et~al.}(2011)}]{Ajith:2009bn}
Ajith, P. {et~al.} 2011, Phys. Rev. Lett., 106, 241101

\bibitem[{Akutsu {et~al.}(2019)}]{Akutsu:2018axf}
Akutsu, T. {et~al.} 2019, Nature Astron., 3, 35

\bibitem[{{Amaro-Seoane} {et~al.}(2017){Amaro-Seoane}, {Audley}, {Babak},
  {Baker}, {Barausse}, {Bender}, {Berti}, {Binetruy}, {Born}, {Bortoluzzi},
  {Camp}, {Caprini}, {Cardoso}, {Colpi}, {Conklin}, {Cornish}, {Cutler},
  {Danzmann}, {Dolesi}, {Ferraioli}, {Ferroni}, {Fitzsimons}, {Gair}, {Gesa
  Bote}, {Giardini}, {Gibert}, {Grimani}, {Halloin}, {Heinzel}, {Hertog},
  {Hewitson}, {Holley-Bockelmann}, {Hollington}, {Hueller}, {Inchauspe},
  {Jetzer}, {Karnesis}, {Killow}, {Klein}, {Klipstein}, {Korsakova}, {Larson},
  {Livas}, {Lloro}, {Man}, {Mance}, {Martino}, {Mateos}, {McKenzie},
  {McWilliams}, {Miller}, {Mueller}, {Nardini}, {Nelemans}, {Nofrarias},
  {Petiteau}, {Pivato}, {Plagnol}, {Porter}, {Reiche}, {Robertson},
  {Robertson}, {Rossi}, {Russano}, {Schutz}, {Sesana}, {Shoemaker}, {Slutsky},
  {Sopuerta}, {Sumner}, {Tamanini}, {Thorpe}, {Troebs}, {Vallisneri},
  {Vecchio}, {Vetrugno}, {Vitale}, {Volonteri}, {Wanner}, {Ward}, {Wass},
  {Weber}, {Ziemer}, \& {Zweifel}}]{2017arXiv170200786A}
{Amaro-Seoane}, P., {Audley}, H., {Babak}, S., {et~al.} 2017, arXiv e-prints,
  arXiv:1702.00786

\bibitem[{Annulli {et~al.}(2020)Annulli, Cardoso, \& Vicente}]{Annulli:2020ilw}
Annulli, L., Cardoso, V., \& Vicente, R. 2020 [\eprint[arXiv]{2007.03700}]

\bibitem[{Barack {et~al.}(2019)}]{Barack:2018yly}
Barack, L. {et~al.} 2019, Class. Quant. Grav., 36, 143001

\bibitem[{Barausse(2007)}]{Barausse:2007ph}
Barausse, E. 2007, Mon.Not.Roy.Astron.Soc., 382, 826

\bibitem[{Barausse {et~al.}(2014)Barausse, Cardoso, \& Pani}]{Barausse:2014tra}
Barausse, E., Cardoso, V., \& Pani, P. 2014, Phys. Rev., D89, 104059

\bibitem[{Barausse {et~al.}(2016)Barausse, Yunes, \&
  Chamberlain}]{Barausse:2016eii}
Barausse, E., Yunes, N., \& Chamberlain, K. 2016, Phys. Rev. Lett., 116, 241104

\bibitem[{Berti {et~al.}(2005)Berti, Buonanno, \& Will}]{Berti:2004bd}
Berti, E., Buonanno, A., \& Will, C.~M. 2005, Phys. Rev., D71, 084025

\bibitem[{Bertone \& Merritt(2005)}]{Bertone:2005hw}
Bertone, G. \& Merritt, D. 2005, Phys. Rev. D, 72, 103502

\bibitem[{Cardoso \& Pani(2019)}]{Cardoso:2019rvt}
Cardoso, V. \& Pani, P. 2019 [\eprint[arXiv]{1904.05363}]

\bibitem[{Carson {et~al.}(2019)Carson, Seymour, \& Yagi}]{Carson:2019fxr}
Carson, Z., Seymour, B.~C., \& Yagi, K. 2019 [\eprint[arXiv]{1907.03897}]

\bibitem[{Eda {et~al.}(2013)Eda, Itoh, Kuroyanagi, \& Silk}]{Eda:2013gg}
Eda, K., Itoh, Y., Kuroyanagi, S., \& Silk, J. 2013, Phys.Rev.Lett., 110,
  221101

\bibitem[{ETWhite(2018)}]{ETWhite}
ETWhite. 2018, \url{https://tds.virgo-gw.eu/?content=3&r=14065}

\bibitem[{Ferrer {et~al.}(2017)Ferrer, da~Rosa, \& Will}]{Ferrer:2017xwm}
Ferrer, F., da~Rosa, A.~M., \& Will, C.~M. 2017, Phys. Rev., D96, 083014

\bibitem[{Flanagan \& Hughes(1998)}]{Flanagan:1997sx}
Flanagan, E.~E. \& Hughes, S.~A. 1998, Phys. Rev., D57, 4535

\bibitem[{Gnocchi {et~al.}(2019)Gnocchi, Maselli, Abdelsalhin, Giacobbo, \&
  Mapelli}]{Gnocchi:2019jzp}
Gnocchi, G., Maselli, A., Abdelsalhin, T., Giacobbo, N., \& Mapelli, M. 2019
  [\eprint[arXiv]{1905.13460}]

\bibitem[{Gondolo \& Silk(1999)}]{Gondolo:1999ef}
Gondolo, P. \& Silk, J. 1999, Phys. Rev. Lett., 83, 1719

\bibitem[{Hild {et~al.}(2011)}]{Hild:2010id}
Hild, S. {et~al.} 2011, Class. Quant. Grav., 28, 094013

\bibitem[{Hui {et~al.}(2017)Hui, Ostriker, Tremaine, \& Witten}]{Hui:2016ltb}
Hui, L., Ostriker, J.~P., Tremaine, S., \& Witten, E. 2017, Phys. Rev. D, 95,
  043541

\bibitem[{Isoyama {et~al.}(2018)Isoyama, Nakano, \& Nakamura}]{Isoyama:2018rjb}
Isoyama, S., Nakano, H., \& Nakamura, T. 2018, PTEP, 2018, 073E01

\bibitem[{Kim \& Kim(2007)}]{Kim:2007zb}
Kim, H. \& Kim, W.-T. 2007, Astrophys.J., 665, 432

\bibitem[{Kocsis {et~al.}(2011)Kocsis, Yunes, \& Loeb}]{Kocsis:2011dr}
Kocsis, B., Yunes, N., \& Loeb, A. 2011, Phys.Rev., D84, 024032

\bibitem[{Krolak \& Schutz(1987)}]{Krolak:1987ofj}
Krolak, A. \& Schutz, B.~F. 1987, Gen. Rel. Grav., 19, 1163

\bibitem[{LIGOWhite(2018)}]{LIGOWhite}
LIGOWhite. 2018, \url{https://dcc.ligo.org/LIGO-T1800044/public}

\bibitem[{Loeb(2016)}]{Loeb:2016fzn}
Loeb, A. 2016, Astrophys. J., 819, L21

\bibitem[{Macedo {et~al.}(2013)Macedo, Pani, Cardoso, \&
  Crispino}]{Macedo:2013qea}
Macedo, C.~F., Pani, P., Cardoso, V., \& Crispino, L.~C. 2013, Astrophys.J.,
  774, 48

\bibitem[{Merritt(2004)}]{Merritt:2003qk}
Merritt, D. 2004, Phys. Rev. Lett., 92, 201304

\bibitem[{Merritt {et~al.}(2002)Merritt, Milosavljevic, Verde, \&
  Jimenez}]{Merritt:2002vj}
Merritt, D., Milosavljevic, M., Verde, L., \& Jimenez, R. 2002, Phys. Rev.
  Lett., 88, 191301

\bibitem[{Navarro {et~al.}(1997)Navarro, Frenk, \& White}]{Navarro:1996gj}
Navarro, J.~F., Frenk, C.~S., \& White, S. D.~M. 1997, Astrophys. J., 490, 493

\bibitem[{Pato {et~al.}(2015)Pato, Iocco, \& Bertone}]{Pato:2015dua}
Pato, M., Iocco, F., \& Bertone, G. 2015, JCAP, 1512, 001

\bibitem[{Peters(1964)}]{Peters:1964zz}
Peters, P. 1964, Phys. Rev., 136, B1224

\bibitem[{Reisswig {et~al.}(2013)Reisswig, Ott, Abdikamalov, Haas, Moesta, \&
  Schnetter}]{Reisswig:2013sqa}
Reisswig, C., Ott, C.~D., Abdikamalov, E., {et~al.} 2013, Phys. Rev. Lett.,
  111, 151101

\bibitem[{Sadeghian {et~al.}(2013)Sadeghian, Ferrer, \&
  Will}]{Sadeghian:2013laa}
Sadeghian, L., Ferrer, F., \& Will, C.~M. 2013, Phys. Rev. D, 88, 063522

\bibitem[{Shapiro~Key \& Cornish(2011)}]{Key:2010tc}
Shapiro~Key, J. \& Cornish, N.~J. 2011, Phys. Rev. D, 83, 083001

\bibitem[{Vallisneri(2008)}]{Vallisneri:2007ev}
Vallisneri, M. 2008, Phys. Rev., D77, 042001

\bibitem[{Vicente {et~al.}(2019)Vicente, Cardoso, \& Zilhão}]{Vicente:2019ilr}
Vicente, R., Cardoso, V., \& Zilhão, M. 2019 [\eprint[arXiv]{1905.06353}]

\bibitem[{Yunes {et~al.}(2011)Yunes, Kocsis, Loeb, \& Haiman}]{Yunes:2011ws}
Yunes, N., Kocsis, B., Loeb, A., \& Haiman, Z. 2011, Phys.Rev.Lett., 107,
  171103

\bibitem[{Yunes \& Pretorius(2009{\natexlab{a}})}]{Yunes:2009hc}
Yunes, N. \& Pretorius, F. 2009{\natexlab{a}}, Phys.Rev., D79, 084043

\bibitem[{Yunes \& Pretorius(2009{\natexlab{b}})}]{Yunes:2009ke}
Yunes, N. \& Pretorius, F. 2009{\natexlab{b}}, Phys.Rev., D80, 122003

\bibitem[{Yunes {et~al.}(2016)Yunes, Yagi, \& Pretorius}]{Yunes:2016jcc}
Yunes, N., Yagi, K., \& Pretorius, F. 2016, Phys. Rev., D94, 084002

\end{thebibliography}

\end{document}